# Stars: Evolution, Stability and Statistical Mechanics[†]

**Anuj Chaudhri**

*Department of Mechanical Engineering and Applied Mechanics, University of Pennsylvania, Philadelphia, PA 19104*

**Abstract**: This paper reviews the physics of stars—the type, structure, evolution and stability. Simple thermodynamics and statistical mechanics are used to show the inner working of white dwarf and neutron stars. The major concentration of the paper will be on white dwarf stars although in some places references will also be made to neutron stars where the relations can be extended easily. It can be shown that a maximum mass limit is attached to each type of star which can be derived rigorously. Maximum entropy can be used to show that the gravitational contraction is balanced by the degeneracy pressure created by the electrons in the case of white dwarfs, and neutrons and protons which constitute the matter of the neutron star. Finally the kinetic equations which describe the luminosity of the star and the radiative transfer are introduced.

## 1. Introduction

Stars are the most beautiful celestial objects in the sky because of the light they emit. They are considered as objects of plasma which produce energy through nuclear fusion in their core. Stars in the main sequence support their luminosity through nuclear reactions in the core which give rise to a radiation pressure. This in turn is supported by the star's immense gravitational pressure (due to its density) and keeps the star in hydrostatic equilibrium during most of its lifetime. The structure, evolution and stability of stars can be well described by simple thermodynamic and statistical physics models which predict a complex process of star evolution till 'death'.

Stars in the main sequence usually burn their nuclear fuel and move towards one of the final three states depending on the mass of the star: white dwarfs, neutron stars and black holes. The more massive stars end up becoming black holes, so called, because of the immense gravitational pressure they possess. These stars are so dense yet so small that anything including light is not allowed to escape from within the star. The subject of black holes is very interesting as they serve as grounds to study both general relativity and quantum mechanics in one framework. However this has led to the breaking down of both theories at the singularity of the black hole and to development of string theory. Stars with mass greater than 3 $M_☉$ where $M_☉$ is the mass of the sun collapse into black holes. Stars with masses between 1.4 $M_☉$ and 3 $M_☉$ collapse into neutron stars. Smaller stars with masses between 0.5 and 1.4 $M_☉$ collapse into white dwarfs. Typically white dwarfs have radii of 5000 km and neutron stars have radii of 10 km.

[†] Title inspired by the Nobel lecture – Dr. Subrahmanyan Chandrasekhar, 1983

In this paper the evolution of stars is introduced first to understand how the mass of the star determines its constituents. This is followed by the understanding the structure and stability of white dwarf stars through the realms of thermodynamics, statistical mechanics and special relativity [1]. Neutron stars are introduced next but a lot of detail is avoided as the study is very complicated due to the addition of general relativity. Finally the kinetic and transport equations are introduced which explain the photon kinetics and radiative transport within the star.

## 2. Stellar Evolution

The study of compact stars began in the early 20$^{th}$ century with pioneering work done by pre-eminent astrophysicists such as Sir Arthur Eddington and Dr. Subrahmanyan Chandrasekhar. It was Dr. Chandrasekhar [2] who first derived equations for the maximum mass that a white dwarf star could have. This limiting mass is called the Chandrasekhar limit in his honor. Later on many physicists added to the body of literature by deriving equations of stability and incorporating effects of general relativity to understand the complexities of stellar evolution.

The evolution of the star before and after the main sequence proceeds in very different ways. The star begins its life as a hydrogen cloud and starts to contract due to self gravitation. As the density and pressure increase, the temperature and luminosity also increase. This protostar then turns into a star if the temperature of its core becomes sufficiently high so that to initiate nuclear reactions of fusion. The emission of light is maintained as long as the fusion reactions take place in the inner cores. During this time the stars are like dynamical objects in which there exists a symbiotic relationship between radiation and matter creating enough pressure to oppose gravitational contraction. Once the nuclear fuel is exhausted, the production of heat in the core ceases and the gravitational contraction continues. For masses between 0.5 M☉ and 1.4 M☉, fusion results in the plasma of $^4$He nuclei and electrons. As long as this gas behaves classically, it contracts rapidly but this stops when the electrons become a Fermi gas. The object becomes a white dwarf, an inert star that shines while cooling down slowly. Its matter can be represented as a neutral mixture of two independent particle gases, the $^4$He nuclei and the electrons, with the electron gas quantum mechanical in nature and the $^4$He gas classical.

As pointed by Dr. Chandrasekhar, stars with masses greater than $M_{ch}$ (Chandrasekhar limit) which is roughly equal to 1.4 M☉, contain electrons which behave ultra-relativistically and hence hydrostatic equilibrium as a cold body is difficult. These objects reach very high temperatures so that the fusion of nuclei produces elements heavier than $^4$He, such as $^{12}$C, $^{16}$O, and $^{20}$Ne. During the collapse of such a star, due to its high temperatures the star explodes ejecting most of the mass into interstellar space, while the core implodes to form a neutron star. During the process the electrons are captured by protons to make neutrons and neutrinos. The neutrinos carry away most of the gravitational binding energy leaving behind the neutron gas to support the star. Hence the neutron star, as the name implies, is composed of a quantum mechanical neutron gas with density comparable to that of nuclear matter.

### 3. White Dwarf Stars

White dwarf stars comprise of the quantum mechanical electron gas and the classical $^4$He gas (referred to as nuclei henceforth). It can be safely assumed that most of the density (mass) of the star is derived from the nuclei, whereas the pressure is derived from the quantum mechanical nature of the electrons, hence electron degeneracy pressure. In general, during the evolution of stars the gravitational collapse is balanced by radiation pressure and matter pressure which in the case of white dwarf stars is due to the degenerate electrons. This can be written as follows,

$$\frac{dP}{dr} = -\frac{GM(r)\rho(r)}{r^2} \tag{1}$$

Where, $P = P_e + P_{rad}$, i.e. the total pressure is the sum of the electron and radiation pressures. Now for white dwarf stars, it can be shown [2] that the radiation pressure at the centre for a star close to solar mass cannot exceed 3 % of the total pressure. Now when gravitational pressure is squeezing on these stars, the electrons in the core are getting squeezed. In the language of quantum mechanics, the electrons get localized and hence due to Heisenberg's uncertainty principle become increasingly frenzied in their motion. This causes a sort of electron pressure which is non-zero even at zero Kelvin due to zero-point motion. Electrons being fermions obey Fermi-Dirac statistics. For a zero-temperature Fermi gas all the available electron states below the Fermi energy are filled. Typical Fermi energies [3] are of ~ 1 MeV, which corresponds to a Fermi temperature of ~ $10^{10}$ K. It is estimated that the temperatures at the interiors of white dwarf stars are of ~ $10^6$ K or so. So even if the electrons want to jump to higher states by an amount of order $k_b T$, they fail. The high energy electrons represent a tiny fraction of the electrons in the star. Hence for computing the pressure of the system it is accurate (1 part in $10^4$) to describe the electrons as a degenerate Fermi gas at zero Kelvin. The radiation pressure becomes important in highly massive stars, but to develop degeneracy, the radiation pressure must not exceed ≈ 10 % of the total pressure. Hence for all purposes, in this paper the pressure is assumed to come from a degenerate Fermi gas at zero Kelvin.

### 3.1 Equation of state of a degenerate electron gas

The equilibrium properties of a gas can be described by the grand potential Ω (T, μ, V), which is a function of the temperature T, chemical potential μ and the volume V. It can be described as,

$$\Omega = -\int d\varepsilon \, \mathrm{N}(\varepsilon) f(\varepsilon) \tag{2}$$

Where, N (ε) is the number of single-particle states with energy less than ε. Now f(ε) is given by,

$$f(\varepsilon) = \frac{1}{e^{(\varepsilon-\mu)/kT} - \eta} \qquad (3)$$

With η = +1 for bosons and η = -1 for fermions. Thus the electrons and neutrons in the stars obey Fermi-Dirac quantum statistics. If D (ε) corresponds to the density of states then we have that,

$$D(\varepsilon) = \frac{dN(\varepsilon)}{d\varepsilon} \qquad (4)$$

$$N(\varepsilon) = \int d\varepsilon D(\varepsilon) f(\varepsilon) \qquad (5)$$

$$U(\varepsilon) = \int d\varepsilon D(\varepsilon) \varepsilon f(\varepsilon) \qquad (6)$$

Where U is the internal energy, N is the particle number.

We can consider two cases, 1) where the classical limit is obtained and η is neglected in (3). This corresponds to when kT >> $\varepsilon_F$ (Fermi energy), and 2) when kT << $\varepsilon_F$, i.e. the quantum limit. The first case is satisfied for the nuclei in the white dwarfs. In some cases the nuclei may crystallize too [1]. In the non-relativistic case i.e. when $mc^2$ >> kT, we have for particles with spin ½,

$$N(\varepsilon_F) = \frac{V}{3\pi^2 \hbar^3} (2m\varepsilon_F)^{3/2} \qquad (7)$$

This expression (7) can be derived independently from density of state calculations also. This is satisfied by the electrons in white dwarfs and neutrons in neutron stars, but not for electrons in heavy white dwarfs where the relativistic limit would hold. Also for fermions the Fermi factor in the quantum limit becomes [1],

$$f(\varepsilon) = \theta(\mu - \varepsilon) - \frac{\pi^2}{6} (kT)^2 \delta'(\varepsilon - \mu) \qquad (8)$$

Equations (6) to (8) give the following relations in the non-relativistic case,

$$\frac{U}{N} = \frac{3}{5}\varepsilon_F = \frac{3}{5}\left(\frac{3\pi^2 N}{V}\right)^{2/3}\frac{\hbar^2}{2m} \tag{9}$$

$$\boxed{U = \frac{3}{5}\left(\frac{3\pi^2}{V}\right)^{2/3}\frac{\hbar^2}{2m}N^{5/3}} \tag{10}$$

Now from $\Omega$ (T, μ, V) = -PV, we can obtain a relation for the pressure,

$$\boxed{P = \frac{2}{3}\frac{U}{V} = \frac{2}{5}\frac{\hbar^2}{2m}(3\pi^2)^{2/3}\left(\frac{N}{V}\right)^{5/3}} \tag{11}$$

In the relativistic case, we will have the following relation for N (ε), U and P,

$$N(\varepsilon_F) = \frac{8\pi V}{3h^3 c^3}\varepsilon_F^3 \tag{12}$$

$$\boxed{U = \frac{3}{5}\left(\frac{3h^3 c^3}{8\pi V}\right)^{1/3}N^{4/3}} \tag{13}$$

$$\boxed{P = \frac{1}{3}\frac{U}{V} = \frac{1}{5}\left(\frac{3h^3 c^3}{8\pi}\right)\left(\frac{N}{V}\right)^{4/3}} \tag{14}$$

Equations (10), (11), (13) and (14) are the basic relations between the pressure, internal energy and density of particles found in stars which contain degenerate electrons in the non-relativistic and relativistic limits respectively.

### 3.2 Stability, mass-radius relations and the Chandrasekhar limit

The stability of a star can be looked at either by looking at the equations of conservation of mass, momentum and energy or by looking at maximum entropy principles. First the conservation equations are used to derive the mass-radius relations and the limiting mass of white dwarf and neutron stars. Equation (1) gives the equation of hydrostatic equilibrium. Another equation that is needed is the equation of mass conservation,

$$\frac{dM(r)}{dr} = 4\pi r^2 \rho \tag{15}$$

M(r) is the mass inside the assumed spherically symmetric star of radius 'r'. This completes the system of equations along with the previously derived equations of state P=P (ρ). The boundary conditions for this system are,

$$M(r) = 0 \big|_{r=0} \tag{16}$$

$$P(r) = 0 \big|_{r=R} \tag{17}$$

In order to understand the limiting mass, it would be necessary to solve these equations and look at the limit of density approaching infinity and the mass approaching a limiting mass, called as the Chandrasekhar limit. Instead of deriving the equations a simple dimensional analysis can be used to understand the implications of this relation [4].

Let P be the pressure inside the star, $\mu_e$ be the 'molecular weight per electron', ρ the density of the star. We can write the number density N/V in terms of the density, mass and 'molecular weight per electron'. A simple analysis gives you the following results from (1), and (15),

$$P \propto \frac{GM^2}{R^4}; \rho \propto \frac{M}{R^3} \tag{18}$$

For low density models, the non-relativistic form of the equation (11) along with (18) lead to the following expressions for the pressure and mass-radius relation,

$$P \propto \frac{P_0}{(\rho_0 \mu_e)^{5/3}} \frac{M^{5/3}}{R^5} \tag{19}$$

$$R \propto \frac{P_0}{G(\rho_0 \mu_e)^{5/3}} M^{-1/3} \qquad (20)$$

Here $P_0$ and $\rho_0$ correspond to physical constants that have been grouped together to form the relations in (19) and (20).

Similar expressions arise when the relativistic equations are used and this leads to the limiting mass equation in equilibrium,

$$M_{ch} \propto \frac{(P_0/G)^{3/2}}{(\rho_0 \mu_e)^2} \propto \frac{(hc)^{3/2}}{G^{3/2} H^2} \mu_e^{-2} \qquad (21)$$

H is the atomic mass unit in (21). Only for $M \equiv M_{ch}$ can gravitational pressure balance the electron pressure. For a slightly larger mass, the gravitational pressure will win and the star will collapse dynamically to a singularity. For a slightly smaller mass, the electron pressure dominates. The star starts to expand, lowering the density and approaching the non-relativistic equation of state in parts of the interior, until equilibrium can be obtained at a finite radius.

For neutron stars similar analysis can be done and based on escape velocities of the neutrons, a limiting mass of neutron stars can be derived at 3 M☉, beyond which it will be difficult for the photons to escape and objects would turn into black holes where light is trapped by gravity [1]. Further due to strong gravitational fields, general relativistic corrections are often used to explain the effects in neutron stars which will not be discussed in this paper. The formation of stars with limiting masses can be understood by looking at figure 1, which clearly outlines the regimes in which white dwarfs and neutron stars originate. This diagram is clearly an equilibrium curve [5] and shows regions of stability and instability.

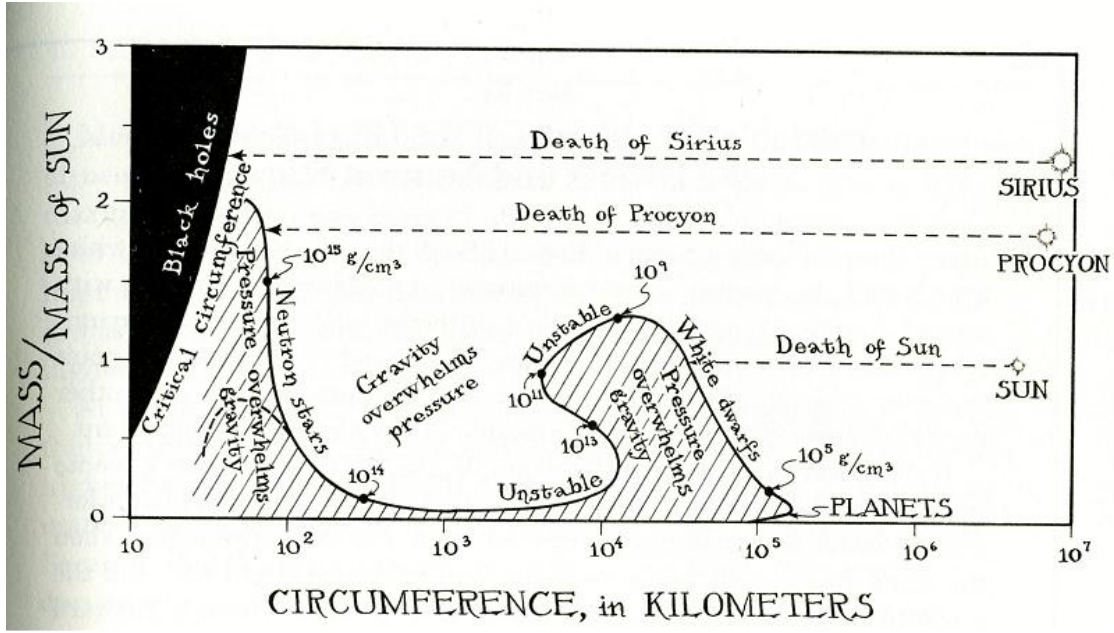

**Figure 1.** Schematic showing the various regions of stable and unstable limits for star configurations [5].

### 3.3 Maximum Entropy considerations

Gravitational equilibrium can also be expressed by looking at the maximum of entropy, subject to constraints on conserved quantities. The existence of radiation within a star causes a negligible departure from equilibrium except for very massive stars. Hence the entropy S is a maximum for a fixed value of energy given by E = U + $E_G$, U being the internal energy and $E_G$ being the gravitational energy. If β is a Lagrange multiplier associated with the constraint of fixed energy, we have to maximize,

$$S - \beta(U + E_G) \tag{22}$$

If $E_G$ is of the form, $E_G = -3GM^2/5R = 3PV$, we have for variations of S and E = 0, the following expression,

$$\frac{dP}{dV}\Big|_S < -\frac{4}{3}\frac{P}{V} \tag{23}$$

Hence we have stable gravitational equilibrium only if the adiabatic compressibility given by (23) is smaller than ¾. This is satisfied for non-relativistic gases, whether classical or quantal, but not for ultra-relativistic Fermi gases. Such a relativistic instability is responsible for the gravitational collapse of massive objects, especially for the implosion of a supernova leading to a neutron star.

**4. Photon Kinetics and Blackbody Radiation**

The main features of light emission can be understood by regarding the surface of a star as a blackbody, i.e. a perfect absorber. The total radiative power given by the luminosity L, emitted by the star can be deduced from the light flux received from it and the distance of the star. The radius R can be obtained from the Stefan-Boltzmann law,

$$L = 4\pi R^2 \sigma T_s^4; \sigma = \frac{\pi^2 k^4}{60\hbar^3 c^2} \qquad (24)$$

$T_s$ is the surface temperature of the star obtained by comparing its spectrum with Planck's law. This luminosity is also used to classify stars, so that the color of a star gives a direct estimate of the surface temperature. The HPR (Hertzsprung-Russell) diagram shown in figure 2 shows the variation of surface temperature and luminosity.

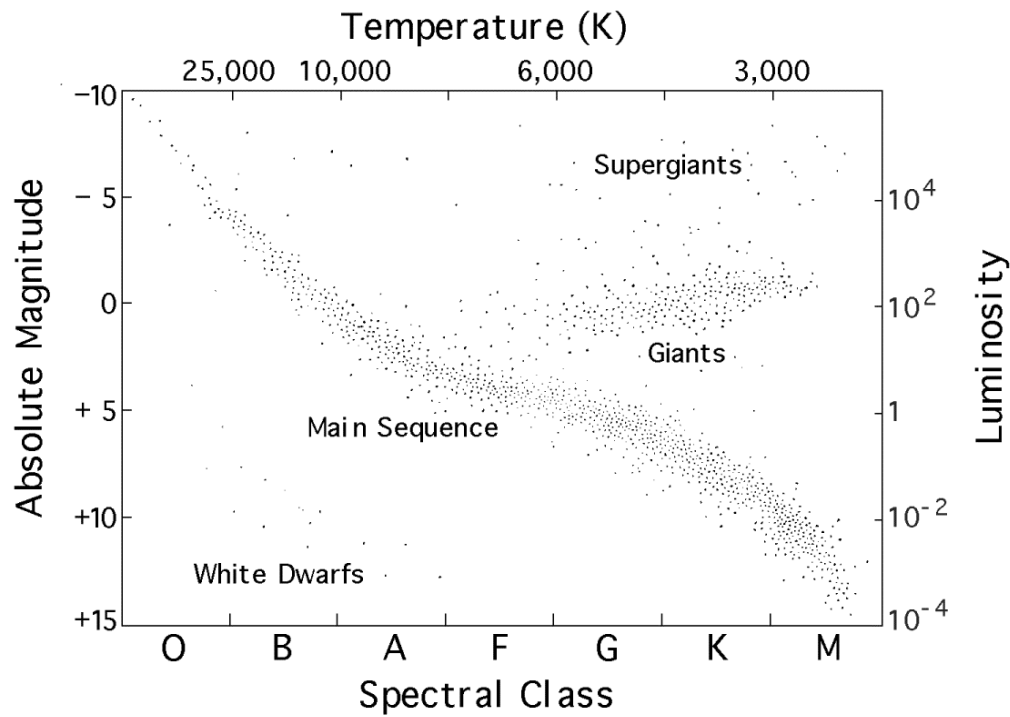

**Figure 2.** Hertzsprung-Russell diagram

(Image source: http://imagine.gsfc.nasa.gov/docs/science/know_l2/stars.html)

To understand this, a more rigorous microscopic theory is needed. The structure of the matter of the star and the ill-defined surface are taken into consideration. The light is assumed to come from a superficial shell which is partly transparent called the photosphere. In this region the temperature T(r) and the density ρ(r) decrease with 'r'. the propagation and thermalization of particles in a gaseous medium, such as photons in the photosphere, are described by the Boltzmann-Lorentz kinetic equation for the particle density f(r, p, t). The equation is as follows,

$$\frac{\partial f}{\partial t} + v \bullet \nabla f = \mathrm{I}(f) \qquad (25)$$

$$\Lambda_\upsilon(r,u,t) = \frac{h^4 \upsilon^3}{c^2} f(r,p,t) \qquad (26)$$

The particle probability density can be changed into a convenient expression in terms of Λ(r, u, t) where u = **p**/p is a unit vector in the direction of propagation. When 'r' lies outside the star, Λ(r, u, t) can be identified as the luminance L. The kinetic equation now looks like,

$$\frac{\partial \Lambda_\upsilon}{\partial t} + cu \bullet \nabla \Lambda_\upsilon = \mathrm{I}_c + \mathrm{I}_{sp} + \mathrm{I}_{st} - \mathrm{I}_a \qquad (27)$$

The term $I_c$ corresponds to the collision term describing elastic or inelastic scattering of photons, $I_{sp}$ and $I_{st}$ are the source terms associated with spontaneous and stimulated emissions and $I_a$ is the absorption term. At each point matter is in thermal equilibrium at temperature T(r), but the photons are far from equilibrium like near the surface where the distribution is extremely anisotropic as they are escaping. In the photosphere, T lies below the ionization temperature so that matter only contains atoms which govern absorption and emission and inside the star where matter is completely ionized the scattering term is the dominant term.

To understand how the luminosity term is measured, the above equation (27) needs to be solved for a given set of absorption and emission terms in the photosphere. If $n_1$ and $n_2$ are two species in thermal equilibrium, the absorption and emission terms can be written as,

$$I_a = c\sigma_a n_1 \Lambda_\upsilon \qquad (28)$$

$$I_{st} = c\sigma_a n_2 \Lambda_\upsilon \qquad (29)$$

Where $\sigma_a$ is the cross-section for absorption of a single photon by a single atom, the flux of photons reaching the atom is 'cf'. The spontaneous emission term $I_{sp}$ can be approximated for each photon mode by dividing the contribution of stimulated emission by the number $N_\gamma$ of photons.

After solving a series of complicated integrals and making approximations of treating the surface of a star as a plane, the solution for $\Lambda(r, u, t)$ can be obtained and from this a relation for $L_\nu(\theta)$, $\theta$ being the direction of propagation and the z-axis. For a complete derivation please look at reference [1]. The luminosity can be written as,

$$L_\upsilon(\theta) = \int_0^\infty \frac{d\zeta}{\cos\theta} e^{-\zeta/\cos\theta} L_\upsilon^0[T(z)]$$

$$L_\upsilon^0(T) = \frac{2h\upsilon^3}{c^2} \frac{1}{e^{h\upsilon/kT}-1} \qquad (30)$$

$\zeta$ is defined as the optical depth which characterizes the penetration of light in the photosphere, which is governed by the cumulative effect on photons of matter lying above z; an incoming beam with $\cos\theta < 0$ would be attenuated as $e^{-\zeta/|\cos\theta|}$ as it propagates inward.

The aim of the entire exercise was to bring to light the fact that luminosity and surface temperature relations can be understood by solving the basic kinetic equations of Non-Equilibrium Statistical Mechanics by assuming local equilibrium at temperature T(r). For a fixed $\theta$ and $\nu$, the luminance relation (30) appears as a weighted superposition of blackbody radiation associated with successive layers. Because the photosphere is thin, its temperature is nearly uniform and so the above treatment is fairly good for the total luminosity and spectrum of emitted light.

## 5. Conclusions

The state of a star at a given time depends on many quantities, mass M, composition which depends on temperature T and density $\rho$, pressure P, energy density u(r), extreme values of temperature $T_s$ and $T_c$ at surface and center of star, gravitation energy $E_G$, internal energy U, luminosity L and the spectral composition of the emitted light. As we have seen in this paper, the mass of the star is the most important quantity and determines the major constituents of the star. The presence of the elements changes with the mass and temperature and influences other properties of the star like luminosity. The equations in this paper assumed that the stars are isotropic, but in general the density and hence the mass of the star can vary and different phenomena could arise. For a given constitution, equilibrium statistical mechanics provides the equation of state P ($\rho$) and relates the pressure, internal energy, density, radius and temperature. The criterion for using quantum mechanics or relativity depends on the density $\rho(r)$ and T(r). The constitution of the matter in the stars determines their evolution. The luminosity of a star is characterized by the photon kinetics in the photosphere which can be calculated by using the Boltzmann-Lorentz kinetic equation. The luminosity is related to the surface temperature which can be used to fit the spectrum of stars based on the light emitted by them.

The cases considered in the paper are very basic and many effects have been ignored. Effects of general relativity, corrections to the equations of state, transport of photons from the interior to the exterior limits of the star are few of the many things that have not been covered. Such effects require additional tools of understanding beyond the scope of this paper. The aim of this paper has been to introduce the fascinating topic of Stellar Thermodynamics and Statistical Mechanics and understand how the equations can be used to derive meaningful relations between diverse quantities and get an intuitive feel for this beautiful subject.